\documentclass[journal]{IEEEtran}
\usepackage[utf8]{inputenc}
\usepackage[T1]{fontenc}
\usepackage{graphicx}
\usepackage{amsmath,amssymb}
\usepackage{siunitx}
\usepackage{url}
\usepackage{xurl}
\usepackage{hyperref}
\usepackage{booktabs}
\usepackage{array}
\usepackage{xcolor}
\usepackage{balance}
\usepackage[final]{changes}
\usepackage[bottom,flushmargin]{footmisc}

\title{Deep Tech to Space: Space Data Centers\\ 
	and AI Revolution at the Edge\small} 

\author{\vspace{-0.2cm}%
	Jonas Weiss$^{1}$, Patricia Sagmeister$^{1}$, Gabriel Maiolini Capez$^{2,3}$, Dinesh Verma$^{4}$, Roberto Garello$^{2}$, Alberto Perotti$^{2}$, Dawid Lazaj$^{5}$, Alicja Musial$^{5}$, Jakub Nalepa$^{6,5,9}$, Thomas Morf$^{1}$, Martin Schmatz$^{1}$, Marek Krawczyk$^{7}$, Mateusz Przeliorz$^{5}$, Kevin Roche$^{4}$, Sagar Tayal$^{8}$, Mahalakshmi Lakshminarayanan$^{8}$, Nicolas Longépé$^{9}$, Pierre-Philippe Mathieu$^{9}$, Agata Wijata$^{6,5}$\\[0.5em]
	\small \smash{$^{1}$} IBM Research Europe, Säumerstrasse 4, 8803 Rüschlikon, Switzerland; \smash{$^{2}$} Politecnico di Torino, Corso Duca Degli Abruzzi, 24, 10129 Torino, Italy; \smash{$^{3}$} Vyoma GmbH, Karl-Theodor-Straße 55, 80803 München, Germany; \smash{$^{4}$} IBM Research, 1101 Kitchawan Rd, Yorktown Heights, NY 10598, United States; \smash{$^{5}$} KP Labs, Bojkowska 37J, 44-100 Gliwice, Poland; \smash{$^{6}$} Silesian University of Technology, ul. Akademicka 2A, 44-100 Gliwice, Poland; \smash{$^{7}$} Meguro Space, Cracow, Małopolskie, Poland; \smash{$^{8}$} IBM, 1 Orchard Rd, Armonk, NY 10504, United States; \smash{$^{9}$} ESA $\Phi$-lab, Via Galileo Galilei, 1, 00044 Frascati RM, Italy.\vspace{-1.1cm}
}

\begin{document}
	\maketitle
	\begin{abstract}
		Dramatic cost reductions driven by private sector innovations have led to a rapid increase in the number of satellites in orbit and a corresponding surge in space-generated data. As this trend continues, transmitting large volumes of data to Earth for processing may become increasingly costly and challenging due to potential space-to-Earth link congestion and increased latency. Moreover, traditional ground station networks may face difficulties accommodating growing data flows and workloads because of capacity constraints, complex scheduling logistics, and restricted visibility windows, which can limit scalability.
		Space data centers (SDCs) -- software-driven, multi-tenant artificial intelligence-based service platforms capable of processing data in orbit to generate actionable insights for client satellites and ground users -- represent a promising approach to address these challenges.
		
		This article presents the architecture of a low Earth orbit SDC satellite constellation, considering orbital design, inter-satellite links and network topology, computational resource organization, and software service orchestration. We analyze the potential technical feasibility and economic viability of SDCs using forecasting models informed by technology roadmaps and illustrate the concept through Earth observation and lunar exploration use cases.
	\end{abstract}
	
	\vspace{-.2cm}

	\vspace{-.5cm}
	\section{Introduction and Motivation}\vspace{-.2cm}

	The rapidly increasing number of satellites in orbit, driven by the recent significant reduction in launch and manufacturing costs, is fueling an unprecedented growth in space-generated data volumes.
	Due to the often limited onboard processing capabilities of satellites, these
	large data volumes must be stored onboard and later transmitted to Earth for
	processing.
	Data transmissions typically occur in bursts during short time intervals wherein satellites experience favorable link quality
	toward the ground station network (GSN) owing to line-of-sight conditions.
	This operational constraint places considerable pressure on dedicated ground infrastructure and introduces delays in data availability.
	
	Currently, the predominant solutions for downloading data from low Earth orbit (LEO) satellites include GSNs and Data Relay Systems (DRS) based on geostationary orbit (GEO) and medium Earth orbit (MEO) satellites.
	These systems rely on radio frequency (RF) and optical (laser) communication technologies. However, both approaches face significant challenges in providing the required flexibility, high throughput, and low-latency communications at affordable cost on a 24/7 basis~\cite{Capez2024}.
	
	\begin{figure*}[thbp]
		\centering\vspace{-.2cm}
		\includegraphics[width=.7\textwidth,trim={2.8cm 5.0cm 2.8cm 4.3cm},clip, height=5.cm]{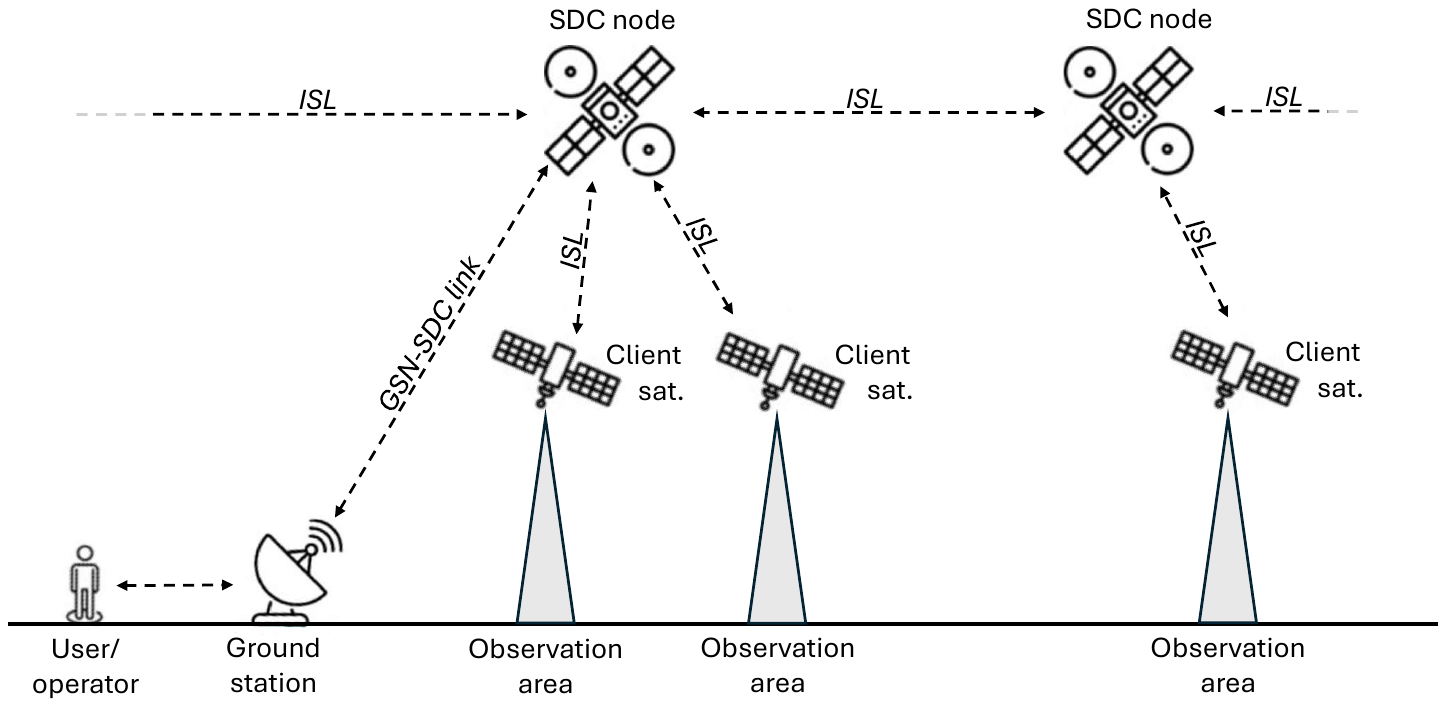}
		\caption{Space Data Center concept: An open inter-satellite network where dedicated data-center satellites provide compute and storage to EO client satellites.}\vspace{-.5cm}
		\label{fig:SDCconcept}
	\end{figure*}
	
	We address these challenges by discussing the emerging concept of Space Data Centers (SDCs) -- software-driven, multi-tenant, artificial intelligence (AI) enabled service platforms capable of processing data directly in orbit.
	By generating low-latency actionable insights for client satellites and ground users, 
	SDCs aim to reduce transmission overhead and minimize delays.
	\added{As a major novelty, this paper introduces a SDC architecture based on \emph{rings of pods}, obtained by identifying the \emph{point of delivery} (pod) as the core building block of a terrestrial data center and reorganizing it within a dedicated LEO constellation structure. This architecture is complemented by a software-centric service organization and a technology-roadmap-based framework for evaluating the SDCs future viability.
	Assessment is conducted using a purpose-built forecasting tool developed specifically for this study and made freely available \cite{Weiss2024}. The tool integrates multiple technology roadmaps to estimate future key performance indicators and associated costs.}
	
	This study was done in order to support the European Space Agency (ESA) efforts
	to explore the feasibility of processing and storing data in orbit to mitigate
	data-transport challenges.\footnote{\url{https://www.esa.int/Enabling_Support/Preparing_for_the_Future/Discovery_and_Preparation/Knowledge_beyond_our_planet_space-based_data_centres}.}

	\vspace{-.3cm}
	\section{Current Space Data Transport Solutions and Their Limits}
	\label{sec:sdtsolutions}
	The primary solutions for downloading large data volumes to Earth are based
	on GSNs and DRSs employing GEO or MEO satellites as relays.
	RF communication remains the predominant technology for space-to-Earth links,
	whereas \added{Inter-Satellite Links (ISLs)} rely on either RF or free-space optical
	(FSO) communication.
	
	Despite their widespread adoption, GSNs exhibit several limitations stemming
	from finite capacity, complex scheduling logistics, and restricted visibility
	time windows.
	Typically, a satellite in Sun-synchronous orbit (SSO) achieves download speeds
	ranging from tens to hundreds of \added{Mb/s}, with visibility windows lasting only a
	few minutes per pass over a ground station.
	Although high download rates may be achievable during these intervals, maintaining
	extended visibility periods -- e.g., several hours per day -- can be prohibitively expensive.
	Furthermore, expanding the number of ground stations to reduce latency and
	improve data download capacity is challenging.
	Certain ground station locations, such as in remote or polar regions, involve
	high deployment and maintenance costs and may suffer from limited terrestrial
	connectivity.
	The addition of new sites also increases the complexity of scheduling and
	frequency allocation.
	
	DRSs, which employ relay satellites in higher orbits, can
	complement GSNs by enabling more continuous data transmission.
	However, DRS deployments are economically challenging, with only a limited 
	number of systems currently available, each comprising a small number of
	dedicated relay spacecraft.
	Moreover, given the large number of satellites concentrated in SSOs at
	altitudes between 500 and 750 km, establishing and scheduling connections
	between many LEO satellites and a limited number of relay satellites is
	highly challenging.
	These systems also face constraints in transmission capacity and service
	availability, due to limited scalability to support a large number of 
	users~\cite{Wang2022}.
	
	RF communication remains the backbone of satellite-to-ground data transmission
	because of its robustness and reliability.
	Different frequency bands, including VHF, UHF, S, X, and Ka, are employed
	depending on mission requirements.
	However, each band is subject to varying levels of path loss, environmental
	effects, interference, and bandwidth limitations, which impact overall
	transmission efficiency and reliability.
	For example, state-of-the-art Earth observation (EO) satellites rely on microwave links with several hundreds of MHz of allocated bandwidth and multi-\added{Gb/s} peak data rates.
	However, increasing spatial and spectral resolutions of onboard sensors demand even higher data rates, achievable only through the larger bandwidths available at higher frequencies, such as the Ka band.
	Operation at these frequencies entails increased path loss, requiring high-gain
	antennas and more precise attitude control.
	Consequently, reliance on GSNs and GEO/MEO relay systems alone does not provide
	a scalable solution for supporting growing satellite constellations, increasing
	data volumes, and stringent latency requirements.
	
	FSO communication provides an alternative to RF links, offering
	significantly higher data rates due to higher carrier frequencies and narrower
	beam widths~\cite{Wang2022,NASA_TBIRD_2022}.
	However, satellite-to-Earth optical communication is highly sensitive to
	atmospheric conditions, particularly cloud cover, which can disrupt laser
	beams and interrupt transmission.
	In addition, the infrastructure required to support optical communication
	is less widespread and more costly, requiring precise alignment with fast-moving
	LEO satellites and making acquisition, pointing, and tracking highly challenging.
	These requirements impose additional operational constraints on spacecraft,
	including increased size, weight, and peak power demands~\cite{Krynitz2021}.
	The limited availability of optical ground stations further exacerbates these
	challenges, reducing system flexibility and increasing vulnerability to geographic
	and atmospheric disruptions.
	
	The limitations of both GSNs and DRS, as well as RF and FSO links highlight
	the need for innovative approaches.
	The inability to guarantee real-time data availability often necessitates
	over-provisioning of onboard computational resources, leading to inefficiencies
	and under-utilization.
	Furthermore, the requirement for extensive onboard storage to compensate for
	communication gaps restricts the achievable survey area and limits the ability
	to dynamically update mission profiles in response to emerging operational needs.
	
	In summary, existing satellite-to-ground data transmission methods, while
	functional, present significant limitations that constrain the efficiency and 
	effectiveness of modern satellite missions.
	Addressing these challenges requires both technological innovation and strategic infrastructure development to enable reliable, low-latency communication capable
	of meeting the growing demand for data exchange.
	These considerations motivate the adoption of advanced solutions that process
	data directly in space, thereby reducing the volume of information transmitted
	to Earth while enhancing satellite autonomy and operational efficiency.

	\vspace{-.4cm}
	\section{Space-Based Data Centers as Softwarized Service Platforms}\vspace{-.1cm}
	\label{sec:SDCconcept}
	A Space Data Center (Fig. \ref{fig:SDCconcept}) is a software-driven
	service platform supported
	by a system of interconnected, cooperating satellites -- the SDC nodes --
	that jointly provide computational, storage, and connectivity services for a
	set of client satellites.
	As the name suggests, the core concept of SDCs closely resembles that of
	terrestrial data centers: each SDC node is envisioned to provide
	functionalities similar to a terrestrial data center's \emph{point of delivery}
	(pod) -- a module of networking, compute, storage, and application components
	that operate together to deliver services and constitute the smallest
	self-contained unit within a data center.
	
	SDCs aim to provide on-demand computational resources for client
	satellites, thereby reducing their hardware footprint and enabling real-time
	data processing directly in orbit.
	By adopting an open architecture with standardized \added{Application Programming Interfaces (APIs)}, SDCs offer flexible,
	multi-tenant access and support a wide range of downstream applications.
	This approach enables scalable AI-driven services while minimizing
	data transfer to Earth.
	Demonstrations by entities such as \emph{Hewlett Packard Enterprise} and the 
	ESA validate the viability of the SDC concept, highlighting
	the potential of these systems to complement terrestrial infrastructure
	and create new economic opportunities in space data processing.
	Companies like Starcloud\footnote{\url{https://www.starcloud.com/}.}
	are developing SDC satellites carrying powerful Graphics
	Processing Units (GPUs).
	However, appropriate design and trade-off analysis of SDCs are critical to
	delivering efficient in-orbit solutions.
	
	A \textit{software-centric} SDC enables straightforward reconfiguration to adapt to
	unforeseen use cases, thereby extending the usability and operational lifetime of satellites.
	In particular, rapid advances in AI~\cite{Wijata2023} 
	outpace traditional satellite design cycles, making static, hardware-specific 
	implementations obsolete by the time of launch.
	A software-centric approach, instead, enables on-the-fly deployment of new
	AI capabilities.
	
	Software-driven design flexibility enables numerous use cases,
	such as updating software-defined radios with new communication standards
	or enhancing satellite image analysis for smart farming and environmental monitoring~\cite{Wijata2023}.
	Additionally, real-time collision avoidance~\cite{GonzaloColombo2021}, anomaly
	detection from satellite telemetry~\cite{Uriot2022}, and in-orbit satellite
	traffic management and maintenance~\cite{Opromolla2024} will rely heavily on
	online system upgrades, as their state of the art is still evolving.
	These opportunities create new economic prospects for satellite operators and
	downstream application providers, particularly when considering constellation-wide
	data aggregation and distributed processing pipelines, which enable scale and
	deep data insights unattainable with the limited resources of individual satellites.
	The tight network interconnection within an SDC constellation further ensures
	continuous ground station connectivity for critical and real-time applications.

	\vspace{-.2cm}
	\section{\added{The Rings-of-Pods Architecture}}\vspace{-.0cm}
	\label{sec:RoP}
	The physical architecture of a data center is tightly coupled with the
	underlying network structure.
	If a single SDC node is considered equivalent to a data center pod,
	the physical architecture is strongly governed by orbital dynamics and
	the network topologies resulting from the orbital parameters.

	In principle, each SDC node could operate on an independent orbit, and
	client satellites could connect to the most suitable SDC node based on
	proximity or link quality.
	Similarly, SDC nodes could interconnect according to the same criteria,
	which, for certain constellations, might result in irregular and non-uniform
	satellite coverage across specific regions.
	To avoid such limitations, we assume that a purpose-built SDC
	constellation consists of different groups of nodes, where nodes within the same
	group share a common orbit and form an SDC node \emph{ring}.
	The constellation would consist of multiple such rings interconnected
	at least near their orbital crossing points.
	This structure enables more predictable network and service availability,
	facilitates uniform coverage and balanced computational workloads, and
	significantly reduces the complexity of dynamically interconnecting nodes
	across different orbits.
	
	\begin{figure}[!tbhp]
		\centering\vspace{-.2cm}
		\includegraphics[width=.98\columnwidth, height=5.7cm]{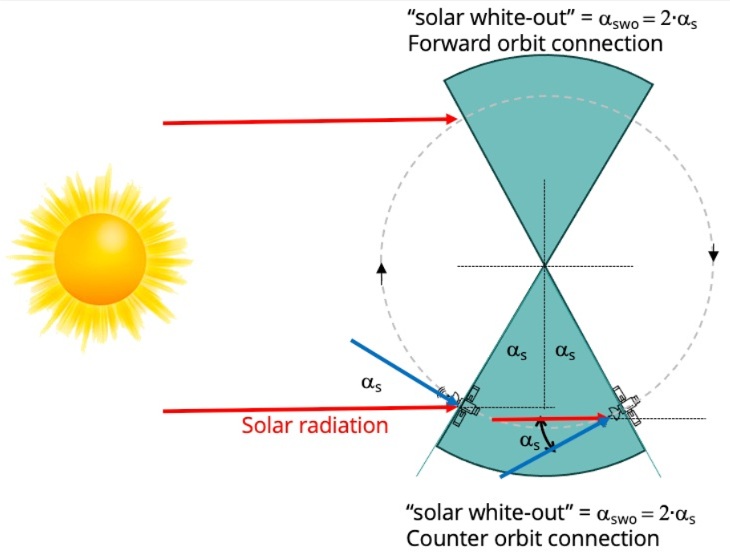}
		\caption{Stray light sensitivity of the optical receiver affects link performance.
			If the receiver faces the sun with an incident angle \added{smaller than} $\alpha_s$, bandwidth will start deteriorating. This will only happen in one direction of the link, either counter or forward orbit-wise, but of course affects link bidirectionality.}\vspace{-.2cm}
		\label{fig:straylight}
	\end{figure}
	
	Under these assumptions, two communication hierarchies of SDC nodes emerge.
	All nodes on a given orbit form a \added{quasi}-static cluster, or \emph{ring of pods}, \added{where each node is statically} interconnected with \added{adjacent} nodes on the same orbit.
	\added{Moreover, each node establishes two} dynamic connections with nodes in other rings.
	
	\added{Aiming to serve clients over most inhabited regions while maintaining a regular and scalable network topology, the widely used Walker-Delta (WD) constellation architecture
	is a strong candidate for the envisioned SDC structure. 
	A WD constellation consists of $P$ orbital planes inclined by $i$ degrees with respect to the equatorial plane.
	The orbital planes are uniformly distributed over the $360^\circ$ range of Right Ascension of the Ascending Node (RAAN), while each plane contains $N_S$ satellites in circular orbits at the same altitude. The constellation is identified by the parameters $(i:N_T,P,F)$, where $N_T=N_S P$ is the total number of satellites and $F$ is the phasing parameter, which specifies the relative phase offset between satellites in adjacent orbital planes. As an example, we consider a constellation with parameters $(i:N_T,P,F) = (53^\circ: 200, 20, 10)$ at an altitude $\Lambda = 600$ km. These values have been selected as a representative reference constellation inspired by existing commercial LEO deployments. They are intended to demonstrate the proposed networking architecture rather than to represent an optimized constellation design.}

	\begin{table*}
	\centering
	\caption{\added{Architectural differences between the proposed architecture and representative alternatives.}}\vspace{-.2cm}
	\label{tab:archComp}\vspace{-.1cm}
	\begin{tabular}{|l|l|l|l|l|}
		\hline 
		\added{\textbf{Architecture}} & \added{\textbf{Processing location}} & \bf \added{Communication requirement} & \bf \added{Resource sharing} & \bf \added{Scalability} \\
		\hline
		\added{Conventional EO architecture}  & \added{Terrestrial data center}	& \added{High downlink (DL) capacity} & \added{High (ground)} & \added{High} \\
		\added{On-board satellite processing} & \added{Individual satellite} & \added{Low DL capacity} & \added{Limited} & \added{Limited by sat. resources} \\ 
		\added{Proposed rings-of-pods SDC}    & \added{Distrib. in-orbit data ctr.} & \added{Moderate DL, high ISL usage} & \added{High across constellation} & \added{High} \\
		\hline
	\end{tabular}\vspace{-.5cm}
	\end{table*}

	To achieve long-distance links (of the order of several thousand \added{km}),
	while ensuring data privacy \footnote{Links between satellites should be considered as
		``private,'' analogous to links between racks in a data center),} and avoid
	congestion of the RF spectrum, \added{highly directional} FSO
	laser links between neighboring satellites are \added{adopted} for intra- and inter-ring connections.
	Equidistant satellite placement along an orbit is not strictly required, but is desirable for practical deployment.
	To estimate benchmark requirements, we assume equidistant satellites.
	
	FSO technology provides the means for very long range links
	virtually not interfering with each other at very large bandwidth, thereby
	overcoming many limitations associated with RF links.
	However, they may be impaired by solar stray light and other phenomena such as cosmic
	radiation and solar flares. While the latter are too sporadic to be modeled
	in this context, solar stray light may severely impact FSO link performance.
	It has been shown that Sun angles as small as $2^\circ$ are achievable for deep
	space missions\footnote{\url{https://ipnpr.jpl.nasa.gov/progress_report/42-154/154L.pdf}};
	however, the SDA OCT Standard T0/T1 explicitly
	excludes any boresight angle (at both receiver and transmitter)
	smaller than $30^\circ$ with respect to the Sun.
	Thus, outside this $30^\circ$ range, no performance impairment is expected,
	whereas terminal behavior within this range is typically unspecified.
	We therefore assume that the \added{ISL} is completely shut down
	for protection when operated within a $30^\circ$ angle from the Sun \added{-- see Fig. \ref{fig:straylight}.}
	A gradual link degradation (e.g., following a Gaussian distribution) would
	also represent a valid assumption, but is not considered here for simplicity.
	Under this assumption, worst-case delays can be \added{determined when data must be sent through different routes}, or caching memory requirements can be
	estimated \added{if no alternative route is available}.
	
	To efficiently exchange data and orchestrate or distribute computationally
	intensive workloads among SDC nodes \added{via ISLs}, the associated routers
	are expected to be more capable than those used to
	communicate with ground stations or acquire data from client satellites.
	We therefore distinguish between \added{three} types of routers:
	\begin{itemize}
		\item \added{External network routers, connecting ground stations or client satellites
		to SDC nodes;}
		\item Dynamic routers, \added{interconnecting SDC nodes on different rings};
		\item Quasi-static routers, interconnecting SDC nodes \added{within each ring}.
	\end{itemize}
	\added{All three types of} routers provide equivalent functionality but are
	optimized for different data volumes and for either dynamic or stable
	connectivity conditions.
	Furthermore, due to the aforementioned FSO link impairments, network routing
	optimization must also consider the direction of data transmission within an
	orbit to avoid inter-satellite links degraded by solar stray light.
	Direct links between client satellites and SDC nodes may be similarly affected,
	depending on their relative positions.
	
	\added{The proposed rings-of-pods architecture naturally supports a hierarchical control and management framework. At the lower level, each orbital ring forms a quasi-static cluster, enabling simple static intra-ring routing, localized resource management, and lightweight network monitoring. At the higher level, neighboring rings are interconnected through dynamic inter-ring optical links, requiring adaptive routing and workload orchestration based on the instantaneous network topology and resource availability. This hierarchical organization reduces control overhead, facilitates scalability, and provides inherent resilience by allowing traffic and computational workloads to be redirected through alternative paths when inter-ring links become temporarily unavailable due to Sun blinding, pointing constraints, or other transient impairments.}
	
	\added{The regularity of the WD topology further simplifies network management by providing predictable connectivity patterns that can be exploited for proactive routing, scheduling, and resource orchestration.}
	
	\added{Data buffering at each terminal may be needed in order to face temporary link outages. The buffering requirements are largely mitigated by the proposed network architecture. The selected WD inclination and phasing produce few Sun-blinding events for intra-ring ISLs, while the 4-ISL satellite architecture and hierarchical routing maintain end-to-end connectivity by dynamically rerouting traffic around temporarily unavailable links. Consequently, local buffering is expected to be required primarily during route convergence or short-lived link impairments rather than throughout the entire duration of a Sun-blinding event. A detailed analysis of buffer sizing, memory bandwidth, and storage technology is left for future work.}

	\added{To complement the architectural discussion, we developed a lightweight WD constellation simulator implementing the proposed ring-of-pods topology and optical ISLs}\footnote{\url{https://github.com/albertogp71/SpaceDataLEO/}}. 
	\added{We conducted a concise evaluation with the orbital configuration described above and setting the ISL range $R_{\rm ISL} = 3000$ km and throughput $T_{\rm ISL} = 100$ Gb/s. Results show an average shortest-path end-to-end latency of 63.05 ms, a worst-case latency of 121.7 ms, and an aggregate network throughput of 72.65 Tb/s for the reference 200-satellite constellation. The average (maximum) number of hops between any two nodes is 7.9 (15.3).
	Although simplified, these results provide a first-order validation that the proposed architecture can support the communication requirements of distributed SDCs. A comprehensive evaluation including physical-layer link budgets, packet loss, congestion dynamics, routing protocol performance, and scalability under realistic traffic patterns is left for future work.}
	
	\added{Table~\ref{tab:archComp} summarizes the main architectural differences between the proposed SDC architecture and representative alternatives. The proposed solution seeks to balance communication efficiency with distributed resource sharing by exploiting optical inter-satellite connectivity.}

	\vspace{-.3cm}
	\section{Use Cases and Quantitative Assessment}	\vspace{-.1cm}
	\label{sec:usecases}
	We present three representative use cases and assess the corresponding
	in-orbit computing requirements. These use cases further explore potential SDC applications beyond the original scope and provide valuable insight
	into the computational demands expected in practical operational scenarios.
	
	\added{The selected use cases are intended to illustrate the architectural motivation for SDCs rather than merely to estimate computational workloads. Modern Earth-observation payloads generate data volumes that increasingly exceed the available space-to-ground communication capacity, making continuous transmission of raw sensor data impractical. AI-enabled in-orbit processing can substantially reduce the volume of data that must be transmitted by extracting higher-level information, detecting events of interest, or generating compressed products directly in space. The proposed distributed SDC architecture extends this concept by enabling computationally demanding AI workloads to be shared across multiple satellites instead of being constrained by the limited resources available on an individual spacecraft. The workload estimates presented in the following use cases therefore serve to assess the feasibility of supporting representative AI-enabled Earth-observation services using future space-based data centers.}
	
	\textit{Use Case 1} involves two satellites on the same orbit, where
	the leading \emph{scout} satellite captures large field of view but coarse resolution Earth observations and sends data to the trailing \emph{mothership} satellite for processing.
	The mothership then acquires detailed scans of identified regions of interest on its own, relaying results at low latency to ground stations. 
	As a compute and networking \textit{workload} example for this use case, we consider wildfire detection \cite{Tuia2022} based on U-Net-based semantic segmentation, which is well documented, scalable and demands low latency \cite{Rashkovetsky2021}. We adopted the Sentinel-2 swath width of 290 km, assumed an 800 km LEO with $\sim$7.5 km/s orbital speed, with $\sim$40 seconds to acquire a 290 $\times$ 290 km$^2$ image. With an assumed 100 m scout resolution and 10 m resolution for the mothership, 3-channel (RGB) 8-bit images, a scout data stream of 630 kBytes/s
	is obtained.
	From \cite{Grabowski2021}, we derive a range of $\sim$272 kFLOP to 1.9 MFLOP of normalized compute-cost per pixel to run a U-Net with 3-channel (RGB) 8-bit per channel input images, identical to $\sim$91 to $\sim$633 GFLOP/MB per input data. With a mean of 362 GFLOP/MB, the average compute needed is thus $\sim$228 GFLOPS for both the scout and the mothership data.
	
	\textit{Use Case 2} involves LEO SDCs gathering data and GEO based SDC
	relay stations with continuous ground connectivity. As for the workload,
	we assume the same segmentation task as Use Case 1, scaled to
	a future application which we assume will produce images of double size (50 MB)
	compared to Use Case 1 and a shorter mean recurrence interval of 20 s.
	The resulting SDC node compute requirement is $\sim$905 GFLOPS for each
	client satellite.
	
	\textit{Use Case 3} applies the SDC concept to a lunar lander SDC and rover client ``constellation,'' with Lagrange point L1 relays replacing GEO relays.
	The workload is generated by mapping resources by means of lunar rovers.
	With very low typical rover speeds -- less than 1 km/h\footnote{\url{https://en.wikipedia.org/wiki/Curiosity}.}, 
	we expect data objects to be small and infrequent, about 3 MB per minute.
	Even with 20 rovers operating concurrently, the total compute requirement for segmentation yields only $\sim$0.3 TFLOPS.
	
	\added{The computational requirements associated with the considered use cases are represented by indicative FLOP estimates derived from the literature and engineering judgment. These estimates are intended to capture the order of magnitude of the required processing effort and to support architectural comparison and technology trend analysis. The actual computational demand depends on the specific implementation, including the AI model, numerical precision, hardware platform, and software optimization. Therefore, the reported values should be interpreted as representative workload estimates rather than validated application benchmarks.}
	\added{While the present work includes a first-order communication evaluation of the proposed constellation architecture, application-level validation of representative Earth observation processing pipelines remains an important topic for future research.}
	
	\begin{figure}[!t]
	\centering
	\includegraphics[width=0.95\columnwidth]{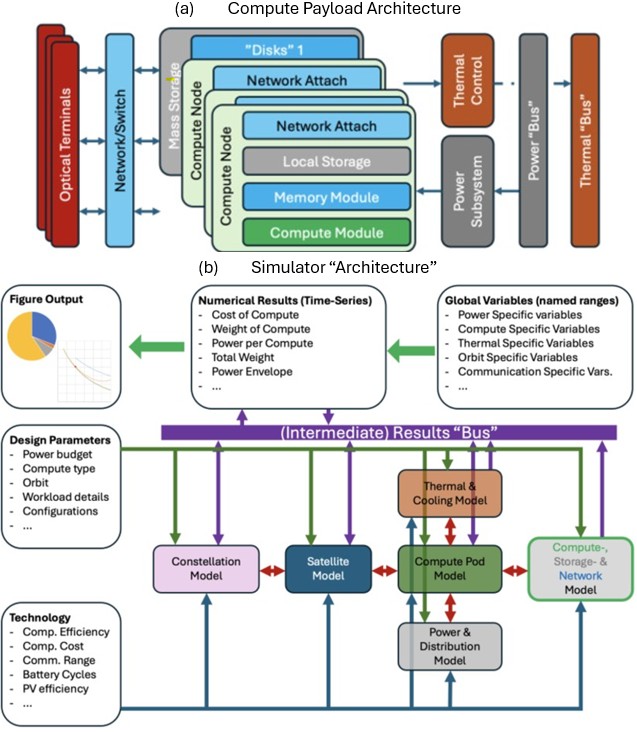}
	\caption{(a) Space Data Centre payload; (b) simulation
		architecture.}\vspace{-.5cm}
	\label{fig:SDCpayload}
	\end{figure}

	\vspace{-.3cm}
	\section{Management, Automation, and Standardization}\vspace{-.2cm}
	\label{sec:management}
	\added{Open standards for space infrastructure, software, APIs, and space-to-space communications are essential to enable interoperable SDC ecosystems. Their widespread adoption will require harmonized international regulations governing spectrum, security, privacy, and commercial operations. In parallel, viable business models are needed to justify the substantial upfront investment required by SDC deployments. Public cloud computing provides a natural reference for developing such models.}
	
	\added{SDCs naturally map onto the NIST cloud computing model \cite{MellGrance2011}, providing on-demand compute, storage, networking, and communication resources through standardized interfaces. Accordingly, conventional cloud service models—including Infrastructure as a Service (IaaS), Platform as a Service (PaaS), and Software as a Service (SaaS)—can be directly translated to the space domain. While IaaS primarily targets large infrastructure providers, PaaS and SaaS lower the entry barrier for smaller organizations and application developers.}
	
	New, emerging public cloud business models are increasingly developed around data and AI.
	The most prevalent data-driven model is \emph{Data as a Service (DaaS)}, where organizations collect, anonymize, and commercialize data from users, either as raw data or processed, through marketplaces\footnote{\url{https://bmilab.com/blog/2022/11/23/the-characteristics-of-data-driven-business-model-development-and-how-to-succeed}.\label{refShaefer}}. 
	\emph{AI as a Service (AIaaS)} enables businesses to incorporate AI into their offerings without in-house expertise\footnote{\url{https://fourweekmba.com/what-is-a-business-model}.}, 
	with \emph{Insights-as-a-Service} (Insights-aaS) providing additionally actionable analytical results across various sectors\footnote{\url{https://newspaceeconomy.ca/2023/06/10/what-are-space-economy-business-models}.}. 
	\emph{Information as a Service (Info-aaS)} utilizes self-collected or third-party data to generate and distribute analytical reports\footref{refShaefer}. 
	\emph{Answers as a Service (AaaS)} provide clients with answers to their questions, necessitating close customer relationships\footref{refShaefer}. 
	All these models rely on the availability of vast amounts of diverse data, storage capabilities, and processing power.
	
	The New Space Economy also provides a fertile ground for \emph{emerging space-based business models}, which are further fueled by a generally observable shift of the entire space industry to ``\textit{as a service}'' \added{models}, where operators move away from owning space infrastructure to leasing it. \emph{Ground Segment as a Service (GSaaS)}\footnote{\url{https://contents.pwc.com/main_trends_challenges_space_sector_22}.\label{refScatteia}}
	or \emph{Ground Station as a Service (GSaaS)}\footnote{\url{https://aws.amazon.com/ground-station/pricing}.}, 
	\emph{Satellite as a Service (SataaS)} and \emph{Payload as a Service (PalaaS)} cater to customers who prefer not to own or operate ground stations or satellites, while \emph{Space Data as a Service (SDaaS)} is rapidly growing\footnote{\url{https://www2.deloitte.com/us/en/insights/industry/aerospace-defense/future-of-space-economy.html}.}, 
	indicating a trend towards the commodification of space data and insights\footref{refScatteia}. 
	With \emph{Resilience as a Service (ResaaS)} as an additional option, identified during one of our study workshops, data and insights generated outside of Earth are ensured to be preserved even in case of system failures. All models aim at reducing cost and accelerate market entry. SDCs further support this \textit{Space as a Service} trend by aggregating and sharing data from various satellites. This will enable complex real time data insight applications and facilitate new ``as a Service'' business models. Space data is also becoming increasingly integral to our everyday life and businesses\footnote{\url{https://www.kratosspace.com/constellations/articles/space-as-a-service-building-an-ecosystem-of-commercial-infrastructure}.\label{refKratosSpace}}.
	However, it is essential to distinguish between long-term data that merits storage and short-term data that may become obsolete shortly after processing. SDCs enable space-based data lakes, providing large storage for raw data, and supporting its delayed use in AI development. Acting as data relays, they relax ground station requirements, streamline mission planning, support in-space processing for immediate insights, improve efficiency and lower data transport cost. Real time data delivery, particularly for time critical incidents, is another significant market opportunity for SDCs.
	
	By 2040, ``\textit{as a Service}'' models altogether are anticipated to dominate the space sector, further attracting new players and establishing ecosystems\footref{refKratosSpace}.
	Table \ref{tab:XaaS} gives an overview of the discussed, established and emerging ``as a Service'' business models we deem suitable for establishing SDC ecosystems.

	\begin{table}[!t]
		\centering
		\caption{``As a Service'' models for SDCs and the New Space Economy.}\vspace{-.2cm}
		\renewcommand{\arraystretch}{1.1}
		\begin{tabular}{|l|p{6cm}|}
			\hline
			\textbf{Abbreviation} & \textbf{Application} \\
			\hline
			\multicolumn{2}{|c|}{%
				\begin{tabular}[t]{@{}c@{}}
					\textbf{``As a Service'' Business Models Adopted from the Public Cloud}
			\end{tabular}} \\
			\hline
			IaaS & Mostly for large enterprises, due to lack of standards. \\
			PaaS, SaaS & Suitable for smaller and non-traditional players. \\
			DaaS & Provides a path to market entry for asset-less new players and startups. \\
			AIaaS & Suitable for organizations with very limited in-house domain expertise. \\
			Insights-aaS & Suitable for players further down the value chain. \\
			Info-aaS & Suitable for players further down the value chain, closer to end user. \\
			AaaS & Suitable for players further down the value chain, very close to end user. \\
			\hline
			\multicolumn{2}{|c|}{\textbf{Emerging ``as a Service'' Business Models for the Space Sector}} \\
			\hline
			GSaaS & Lowers entry barrier for small asset/satellite owners into ecosystem. \\
			SataaS, PalaaS & Mostly for very specialized organizations with dedicated needs. \\
			SDaaS & For expanding business of existing players in the ecosystem as well as for new entrants. \\
			ResaaS & For established players, expanding into new customer requirements. \\
			\hline
		\end{tabular}\vspace{-.5cm}
		\label{tab:XaaS}
	\end{table}

	A major goal of our investigation is to make well-founded predictions about
	the future technical feasibility and corresponding costs for each of the use
	cases mentioned above based on future technology roadmaps.
	With no open tool available to combine long-term technology roadmaps with space system
	and data center preliminary design capabilities, we developed an interactive MS Excel
	based forecasting tool which is publicly available \cite{Weiss2024}.
	It uses predefined or interactively set design parameters, such as orbit, power
	envelope, and over 50 system technology roadmaps (e.g., compute power efficiency,
	battery specific energy, launch costs) to compute figures like cost per tera-operations
	per second (TOPS), power per TOPS, and SDC power cost.
	\added{The roadmap parameters are derived from publicly available technology trends and industry projections, with the option to modify the underlying assumptions for sensitivity analyses.}
	\added{The tool combines these technology roadmaps with a system-level sizing model to estimate figures of merit such as compute capability, power consumption, and subsystem costs for different technology scenarios. The system model itself is deterministic; consequently, the results are governed by the selected roadmap assumptions. As satellite architectures, mission requirements, and technology choices vary considerably, there is no single historical reference against which the model can be validated. Instead, the tool provides a consistent framework for comparing alternative technology roadmaps, assessing their system-level implications, and identifying trends and technology intersection points.}
	
	\added{SDCs may be realized either as large centralized platforms\protect\footnote{\url{https://ascend-horizon.eu/}} or as distributed constellations of interconnected nodes. While larger systems benefit from economies of scale, distributed architectures improve redundancy and reduce distance to clients. Fig.~\ref{fig:SDCpayload} illustrates a representative modular compute payload and the corresponding simulation model adopted in this work.}	
	
	\begin{table}[!t]
		\centering
		\caption{Key design decisions and resulting figures of merit.}\vspace{-.2cm}
		\renewcommand{\arraystretch}{1.1}
		\label{tab:usecases}
		\resizebox{\columnwidth}{!}{%
			\begin{tabular}{|l|c|c|c|}
				\hline
				\textbf{Design} & \multicolumn{3}{|c|}{\textbf{Use case}} \\ 
				\textbf{Decisions} & \textbf{1} & \textbf{2} & \textbf{3} \\ \hline
				Year on the Roadmap & 2032 & 2032 & 2040 \\ \hline
				Provided Total Power [W] & 500 & 2000 & 300 \\ \hline
				Selected Compute Type & \multicolumn{3}{|c|}{GPU-equivalent} \\ \hline
				\multicolumn{4}{|c|}{\textbf{Results, Figures of Merit}}  \\ \hline
				Comp. power [TFLOPS]$^{(1)}$ & 1.14\hspace{1pt}(0.23) & 3.95\hspace{1pt}(3.6$^{(2)})$ & 14.5\hspace{1pt}(0.3) \\ \hline
				Satellite Weight [kg] & 16 & 63 & 68 \\ \hline
				Comp. Efficiency [W/TFLOPS] & 0.44 & 0.5 & 0.02 \\ \hline
				Cost of Power [Euro/W] & 99 & 97 & 21606 \\ \hline
				Cost of Comp. [Euro/TFLOPS] & 43504 & 49276 & 447000 \\ \hline
		\end{tabular}}
		\vspace{-5pt}
		\begin{flushleft}
			\footnotesize $^{(1)}$ Available compute power within power envelope; numbers in parentheses are use cases' minimum requirement.\\
			\footnotesize $^{(2)}$ Assuming four client satellites connected to each SDC node.
		\end{flushleft}\vspace{-.7cm}
	\end{table}
	
	To compare the \added{use cases} workloads, countless additional assumptions are required, mostly substantiated within the “Technology Roadmaps” section in the simulator. Table~\ref{tab:usecases} reports, in the top three rows after headers, the key SDC design parameters: power availability, year on the roadmap and type of compute technology.
	Together, \added{these} govern available compute, communication, and overall satellite size. As the simulator always fully exploits the available power, the resulting computational capability may exceed the workloads, as indicated in parenthesis.
	
	Generally, the tool facilitates powerful and complex comparison between pre-defined or custom scenarios. While results depend on input quality and many assumptions, we found that it produces meaningful results, helping to estimate system costs and evaluate the SDC concept’s commercial viability.
	
	Historical investments in space have shown significant socioeconomic impact\footnote{\url{https://www.oecd.org/en/publications/the-space-economy-in-figures_c5996201-en.html}.}, 
	while studies on deploying cloud technology reveal revenue growth and cost savings, with an average return of investment four times higher than for on-premise solutions\footnote{\url{https://nucleusresearch.com/research/single/cloud-delivers-4-01-times-the-roi-as-on-premises/}.}.
	These historic trends underline the potential of SDCs as a driver for the New Space Economy and broader market segments, to trigger a renewed, massive and positive impact for economy and society.

	\vspace{-.3cm}
	\section{Open Challenges and Conclusions}\vspace{-.1cm}
	\label{sec:openchallenges}
	With the number of satellites in orbit projected to reach several tens of thousands by
	2040 to 2050, the aggregated data acquisition volume in space is expected to grow accordingly.
	This growth will enable new applications, many of which will benefit from or even require near
	real time insights both on the ground and in space. However, relaying terabytes or petabytes of
	data per second to terrestrial data centers for processing presents substantial challenges.
	These challenges include communication energy overhead, data transfer capacity bottlenecks,
	and also importantly, latency between data acquisition and result availability.
	In this paper, we examined existing challenges related to data transport and
	processing at the edge in space.
	
	We foresee that distributed but interconnected satellite data centers
	offer a potential solution to many of these challenges by adapting technologies, architectural schemes, and business models from terrestrial data centers.
	Using our open source space data center forecasting tool,
	we validated key assumptions and compared system metrics for various use
	cases defined in an ESA-funded study that supported this work.
	Additionally, through workshops with key stakeholders from the new space economy, we
	aligned our assumptions and confirmed the value proposition of space data centers.
	
	\added{We conclude that space data centers are technically feasible today, and their
	capabilities are expected to improve as compute, storage, communication, and launch
	technologies continue to evolve.
	Although commercial deployment will require substantial upfront investment, recent
	studies indicate that space-based data centers are not yet cost-competitive with
	conventional terrestrial data centers under current economic conditions.
	Nevertheless, they represent a promising approach for applications in which data
	are generated in space, such as Earth observation, because in-orbit processing can
	reduce the volume of data transmitted to Earth, lower latency, and alleviate demands
	on space-to-ground communications and ground infrastructure.
	Future reductions in launch costs, together with increasing constraints on terrestrial
	power and communication infrastructure, may further improve the economic attractiveness
	of SDCs.
	A comprehensive end-to-end techno-economic comparison between space-based and terrestrial architectures, including total cost of ownership and service revenue models, is an important direction for future work.}

	\vspace{-.3cm}

\bibliographystyle{IEEEtran}
\bibliography{refs3}

@article{Wijata2023,
  author       = {Wijata, A. M. and others},
  title        = {{Taking Artificial Intelligence Into Space Through Objective Selection of
Hyperspectral Earth Observation Applications: To bring the “brain” close to the “eyes” of
satellite missions}},
  journal      = {IEEE Geoscience and Remote Sensing Magazine},
  year         = {2023}
}

@misc{Weiss2024,
  author       = {Weiss, J. and others},
  title        = {{Space Data Centre (IBM Research Europe)}},
  year         = {2024},
  url          = {https://github.com/IBM/space-data-centre},
  note         = {Accessed 2025-11-10}
}

@ARTICLE{Capez2024,
	author={Maiolini Capez, Gabriel and Cáceres, Mauricio A. and Armellin, Roberto and Bridges, Chris P. and Fraire, Juan A. and Frey, Stefan and Garello, Roberto},
	journal={IEEE Journal on Selected Areas in Communications}, 
	title={On the Use of Mega Constellation Services in Space: Integrating {LEO} Platforms Into {6G} Non-Terrestrial Networks}, 
	year={2024},
	volume={42},
	number={12},
	pages={3490-3504},
	doi={10.1109/JSAC.2024.3459078}}

@ARTICLE{Wang2022,
author={Wang, Peng and Li, Hongyan and Chen, Binbin and Zhang, Shun},
journal={IEEE Transactions on Wireless Communications}, 
title={Enhancing {Earth} Observation Throughput Using Inter-Satellite Communication}, 
year={2022},
volume={21},
number={10},
pages={7990-8006},
doi={10.1109/TWC.2022.3163389}}

@misc{NASA_TBIRD_2022,
  author        = {{Center Goddard Space Flight}},
  title         = {{TBIRD TERABYTE INFRARED DELIVERY}},
  howpublished  = {\url{https://www.nasa.gov/wp-content/uploads/2017/10/tbird_fact_sheet_v2.pdf}},
  note          = {Retrieved 11 14, 2024},
  year          = {2022}
}

@inproceedings{Krynitz2021,
  author    = {Krynitz, K. and Heeseb, C. and Knopp, M. and Schulz, K.-J. and Henniger, H.},
  title     = {{The European Optical Communications Network}},
  booktitle = {16th International Conference on Space Operations},
  address   = {Cape Town, South Africa},
  year      = {2021}
}

@inproceedings{GonzaloColombo2021,
  author    = {Gonzalo, J. and Colombo, C.},
  title     = {{On-board collision avoidance applications based on machine learning}},
  booktitle = {8th European Conference on Space Debris},
  year      = {2021}
}

@article{Opromolla2024,
  author  = {Opromolla, R.},
  title   = {{Future in-orbit servicing operations in the space traffic management context}},
  journal = {Acta Astronautica},
  year    = {2024},
  volume  = {220},
  pages   = {469--477}
}

@misc{Uriot2022,
  author        = {Uriot, T. and Izzo, D. and Sim{\~o}es, L. F. and Abay, R. and Einecke, N. and Rebhan, S. and Merz, K.},
  title         = {{Spacecraft collision avoidance challenge: Design and results of a machine learning competition}},
  note          = {Astrodynamics, 6(2), 121--140, 2022}
}

@article{Tuia2022,
  author = {Tuia, D.},
  year = {2022},
  title = {{Perspectives in machine learning for wildlife conservation}},
  journal = {Nature
Communications}
}

@inproceedings{Rashkovetsky2021,
  author = {Rashkovetsky, D. and Mauracher, L. and Langer, S. and Schmitt, M.},
  year = {2021},
  title = {{Wildfire Detection From
Multisensor Satellite Imagery Using Deep Semantic Segmentation}},
  booktitle = {IEEE Journal of Selected
Topics in Applied Earth Observations and Remote Sensing}  
}

@article{Grabowski2021,
  author = {Grabowski, B. and Ziaja, B. and Kawulok, M. and Nalepa, J.},
  year = {2021},
  title = {{Towards Robust Cloud Detection in
Satellite Images Using U-Nets.}},
  journal = {IEEE International Geoscience and Remote Sensing Symposium
IGARSS}
}

@techreport{MellGrance2011,
  author = {Mell, Peter and Grance, Timothy},
  year = {2011},
  title = {{The NIST Definition of Cloud Computing}},
  institution = {NIST},
  number = {Special Publication 800-145}
}

\end{document}